\documentclass[11pt]{article}
\usepackage{moriond,epsfig,graphics}

\newcommand{\cerenkov}{\v{C}erenkov}
\newcommand{\Kstar}{\overline{K}^{\;\!*0}}
\newcommand{\eg}{e.g.}

\newcommand{\etal}{{\em et al.}}
\def\Journal#1#2#3#4{{#1} {\bf #2}, #3 (#4)}

\def\NIM{\em Nucl. Instrum. Methods}
\def\NIMA{{\em Nucl. Instrum. Methods} A}

\def\PLB{{\em Phys. Lett.}  B}
\def\PRL{\em Phys. Rev. Lett.}
\def\PRD{{\em Phys. Rev.} D}

\def\CJP{\em Chin. J. Phys. (Taipei)}
\def\MPLA{{\em Mod. Phys. Lett.} A}
\def\EPJC{{\em Eur. Phys. J.} C}
\def\EPJD{{\em Eur. Phys. J. direct} C}
\def\CPC{\em Comp. Phys. Comm.}
\def\IEEETNS{\em IEEE Trans. Nucl. Sci.}
\def\APPB{{\em Acta Phys. Polon.} B}

\def\be{\begin{equation}}
\def\ee{\end{equation}}
\def\bea{\begin{eqnarray}}
\def\eea{\end{eqnarray}}

\begin{document}
\begin{table}[t]
\begin{center}
\tabcolsep=10.8mm
\begin{tabular}{ll} 
XXXVI$^{th}$ Rencontre de Moriond &  UMS/HEP/2001-027 \\ 
Electroweak Interactions and Unified Theories  & FERMILAB-Conf-01/068-E \\
Les Arcs, France (10--17 March 2001)  & \\ 
\end{tabular}
\end{center}
\end{table}
\vskip 16pt
\vspace*{10mm}
\title{RARE AND FORBIDDEN DECAYS OF $D$ MESONS}

\author{D.~A.~Sanders,$^1$
E.~M.~Aitala,$^1$
S.~Amato,$^2$
J.~C.~Anjos,$^2$
J.~A.~Appel,$^6$
D.~Ashery,$^{14}$
S.~Banerjee,$^6$
I.~Bediaga,$^2$
G.~Blaylock,$^9$
S.~B.~Bracker,$^{15}$
P.~R.~Burchat,$^{13}$
R.~A.~Burnstein,$^7$
T.~Carter,$^6$
H.~S.~Carvalho,$^2$
N.~K.~Copty,$^{12}$
L.~M.~Cremaldi,$^1$
C.~Darling,$^{18}$
K.~Denisenko,$^6$
S.~Devmal,$^4$
A.~Fernandez,$^{11}$
G.~F.~Fox,$^{12}$
P.~Gagnon,$^3$
C.~Gobel,$^2$
K.~Gounder,$^1$
A.~M.~Halling,$^6$
G.~Herrera,$^5$
G.~Hurvits,$^{14}$
C.~James,$^6$
P.~A.~Kasper,$^7$
S.~Kwan,$^6$
D.~C.~Langs,$^{12}$
J.~Leslie,$^3$
B.~Lundberg,$^6$
J.~Magnin,$^2$
S.~MayTal-Beck,$^{14}$
B.~Meadows,$^4$
J.~R.~T.\ de Mello Neto,$^2$
D.~Mihalcea,$^8$
R.~H.~Milburn,$^{16}$
J.~M.~de~Miranda,$^2$
A.~Napier,$^{16}$
A.~Nguyen,$^8$
A.~B.~d'Oliveira,$^{4,\,11}$
K.~O'Shaughnessy,$^3$
K.~C.~Peng,$^7$
L.~P.~Perera,$^4$
M.~V.~Purohit,$^{12}$
B.~Quinn,$^1$
S.~Radeztsky,$^{17}$
A.~Rafatian,$^1$
N.~W.~Reay,$^8$
J.~J.~Reidy,$^1$
A.~C.~dos Reis,$^2$
H.~A.~Rubin,$^7$
A.~K.~S.~Santha,$^4$
A.~F.~S.~Santoro,$^2$
A.~J.~Schwartz,$^4$
M.~Sheaff,$^{5,\,17}$
R.~A.~Sidwell,$^8$
A.~J.~Slaughter,$^{18}$
M.~D.~Sokoloff,$^4$
J.~Solano,$^2$
N.~R.~Stanton,$^8$
R.~J.~Stefanski,$^6$
K.~Stenson,$^{17}$  
D.~J.~Summers,$^1$
S.~Takach,$^{18}$
K.~Thorne,$^6$
A.~K.~Tripathi,$^8$
S.~Watanabe,$^{17}$
R.~Weiss-Babai,$^{14}$
J.~Wiener,$^{10}$
N.~Witchey,$^8$
E.~Wolin,$^{18}$
S.~M.~Yang,$^8$
D.~Yi,$^1$
S.~Yoshida,$^8$
R.~Zaliznyak,$^{13}$ and
C.~Zhang$^8$\\
\begin{center} (Fermilab E791 Collaboration) \end{center}
}

\address{
\begin{table}[h]
\begin{center}
\tabcolsep=4.0pt
\begin{tabular}{ll} 
\em \small $^1$Univ.\ of Mississippi, Oxford, MS 38677, USA &
\em \small $^2$CBPF, Rio de Janeiro, Brazil \\
\em \small $^3$Univ.\ of California, Santa Cruz, CA 95064, USA &
\em \small $^4$Univ.\ of Cincinnati, Cincinnati, OH 45221, USA \\
\em \small $^5$CINVESTAV, 07000 Mexico City, DF Mexico &
\em \small $^6$Fermilab, Batavia, IL 60510, USA \\
\em \small $^7$Illinois Institute of Tech., Chicago, IL 60616, USA &
\em \small $^8$Kansas State Univ., Manhattan, KS 66506, USA \\
\em \small $^9$Univ.\ of Massachusetts, Amherst, MA 01003, USA &
\em \small $^{10}$Princeton University, Princeton, NJ 08544, USA \\
\em \small $^{11}$Universidad Autonoma de Puebla, Puebla, Mexico &
\em \small $^{12}$Univ.\ of South Carolina, Columbia, SC 29208, USA \\
\em \small $^{13}$Stanford University, Stanford, CA 94305, USA &
\em \small $^{14}$Tel Aviv University, Tel Aviv 69978, Israel \\
\em \small $^{15}$Box 1290, Enderby, BC \ V0E 1V0, Canada &
\em \small $^{16}$Tufts University, Medford, MA 02155, USA \\
\em \small $^{17}$Univ.\ of Wisconsin, Madison, WI 53706, USA &
\em \small $^{18}$Yale University, New Haven, CT 06511, USA \\
\end{tabular}
\end{center}
\end{table}
}
%
\vspace*{121.3mm}
\leftline{\hspace{66.2mm}
\resizebox{35mm}{!}{\includegraphics{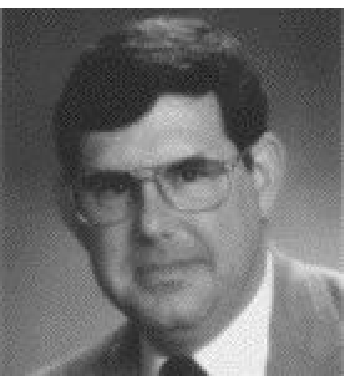}}}
\vspace*{-161.3mm}
%
\maketitle\abstracts{
We summarize the results of two recent searches for flavor-changing 
neutral current, lepton-flavor violating, and lepton-number violating 
decays of $D^+$, $D_{s}^{+}$, and $D^0$ mesons (and their 
antiparticles) into modes containing muons and electrons. Using data 
from Fermilab charm hadroproduction experiment E791, we examined 
$D^+$ and $D_{s}^{+}$ $\pi \ell \ell$ and $K\ell \ell$ decay modes 
and the $D^0$ dilepton decay modes containing either $\ell^+ \ell^-$, 
a $\rho ^0$, $\Kstar$, or $\phi$ vector meson, or a non-resonant 
$\pi \pi$, $K\pi $, or $KK$ pair of pseudoscalar mesons. 
No evidence for any of these decays was found. 
Therefore, we presented branching-fraction upper limits at 90$\%$ 
confidence level for the 51 decay modes examined. Twenty-six of these 
modes had no previously reported limits, and eighteen of the remainder 
were reported with significant improvements over previously published 
results.
}
\newpage
\section{Introduction}
\vskip -10pt
The E791 Collaboration has previously reported limits on rare 
and forbidden dilepton decays of charged charm mesons 
\cite{FCNC,FCNCnew}. Such measurements probe the SU(2)$\times $U(1) 
Standard Model of electroweak interactions in search of new mediators 
and couplings \cite{Pakvasa,SCHWARTZ93}. Here we summarize the results 
of two related analyses. First \cite{FCNCnew} we examined the 
$\pi \ell \ell$ and $K\ell \ell$ decay modes of $D^+$ and $D_{s}^{+}$ 
and the $\ell^+ \ell^-$ decay modes of $D^0$. Then we extend the 
methodology to 27 dilepton decay modes of the $D^0$ meson \cite{4prong} 
containing either resonant $V\ell^+\ell^-$ decays, where $V$ is a 
$\rho ^0$, $\Kstar$, or $\phi$, and non-resonant $h_{1}h_{2}\ell \ell $ 
decays, where $h_{i}$ is either a $\pi $ or a $K$. The leptons were either 
muons or electrons. Charge-conjugate modes are implied. The modes are lepton 
flavor-violating (\eg, \ $D^{+}\rightarrow \pi ^{+}\mu ^{+}e^{-}$), or 
lepton number-violating 
(\eg, \ $D_{s}^{+}\rightarrow \pi ^{-}\mu ^{+}\mu ^{+}$), or 
flavor-changing neutral current decays 
(\eg, \ $D^{0}\rightarrow \Kstar e^{+}e^{-}$). Box diagrams can mimic 
FCNC decays, but only at the $10^{-10}$ to $10^{-9}$ 
level \cite{SCHWARTZ93,Fajfer}. Long range effects through resonant 
modes (\eg, \ $D^{0}\rightarrow \Kstar \rho ^{0}, \ 
\rho^{0}\rightarrow  e^{+}e^{-}$) can occur at the $10^{-6}$ 
level \cite{Fajfer,Singer}. Numerous experiments have studied rare 
decays of charge -1/3 strange quarks. Charge 2/3 charm quarks are 
interesting because they may exhibit a different coupling \cite{Castro}.

The data come from measurements made with the Fermilab E791 
spectrometer \cite{e791spect}. A total of $2 \times 10^{10}$ events 
were taken with a loose transverse energy requirement. These events 
were produced by a 500 GeV/$c$~ $\pi ^{-}$ beam interacting in a fixed 
target consisting of five thin, well-separated foils. Track and vertex 
information came from ``hits'' in 23 silicon microstrip planes and 45 
wire chamber planes. This information and the bending provided by two 
dipole magnets were used for momentum analysis of charged particles. 
Kaon identification was carried out by two multi-cell \cerenkov{} 
counters \cite{Bartlett} that provided $\pi /K$ separation in the 
momentum range $6-60$~GeV/$c$. We required that the momentum-dependent 
light yield in the \cerenkov{} counters be consistent for kaon-candidate 
tracks, except for those in decays with $\phi \rightarrow  K^{+}K^{-}$, 
where the narrow mass window for the $\phi$ decay provided sufficient 
kaon identification (ID). 

Electron ID was based on transverse shower shape plus matching wire 
chamber tracks to shower positions and energies in 
an electromagnetic calorimeter \cite{SLIC}.
The electron ID efficiency varied from 62$\%$ below 9 GeV/$c$ to 45$\%$ 
above 20 GeV/$c$. The probability to misidentify a pion as an electron 
was $\sim 0.8\%$, independent of pion momentum.

Muon ID was obtained from two planes of scintillation 
counters. The first plane (5.5 m $\times$ 3.0 m) of 15 counters 
measured the horizontal position while the second plane (3.0 m $\times$ 
2.2 m) of 16 counters measured the vertical position. There were 
about 15 interaction lengths of shielding upstream of the counters to 
filter out hadrons. 
Data from $D^+\rightarrow \overline{K}^{*0} \mu^{+}\nu _{\!_{\mu}}$ 
decays \cite{Chong} were used to choose selection criteria for muon 
candidates. 
Timing information from the smaller set of muon scintillation counters 
was used to improve the horizontal position resolution. Counter 
efficiencies, measured using muons originating from the primary target, 
were found to be $(99\pm 1)\%$ for the smaller counters and 
$(69\pm 3)\%$ for the larger counters. The probability of 
misidentifying a pion as a muon decreased with increasing momentum, 
from about 6$\%$ at 8 GeV/$c$ to $1.3\%$ above 20 GeV/$c$.
                                                                       
Events with evidence of well-separated production (primary) and decay 
(secondary) vertices were selected to separate charm candidates from 
background. Secondary vertices were required to be separated from the 
primary vertex by greater than $20\,\sigma_{_{\!L}}$ for $D^+$ decays 
and greater than $12\,\sigma_{_{\!L}}$ for $D^0$ and $D_{s}^{+}$ 
decays, where $\sigma_{_{\!L}}$ is the calculated resolution of the 
measured longitudinal separation. Also, the secondary vertex had to be 
separated from the closest material in the target foils by greater than 
$5\,\sigma_{_{\!L}}^{\prime }$, where $\sigma_{_{\!L}}^{\prime }$ is 
the uncertainty in this separation. The vector sum of the momenta 
from secondary vertex tracks was required to pass within 
$40~\mu$m of the primary vertex in the plane perpendicular to the beam. 
The net momentum of the charm candidate transverse to the line 
connecting the production and decay vertices had to be 
less than 300 MeV/$c$ for $D^0$ candidates,
less than 250 MeV/$c$ for $D_{s}^{+}$ candidates, and
less than 200 MeV/$c$ for $D^+$ candidates. 
Finally, decay track candidates were required to pass 
approximately 10 times closer to the secondary vertex than to the 
primary vertex. These selection criteria and kaon identification 
requirements were the same for both the search mode and for its 
normalization signal (discussed below).

To determine our selection criteria, we used a ``blind'' analysis 
technique. Before the selection criteria were finalized, all events 
having masses within a window $\Delta M_S$ around the mass of the 
$D^{0}$ were ``masked'' so that the presence or absence of any 
potential signal candidates would not bias our choice of selection 
criteria. All criteria were then chosen by studying events generated 
by a Monte Carlo (MC) simulation program \cite{MC} and background 
events, outside the signal windows, from real data. The criteria were 
chosen to maximize the ratio $N_{MC}/\sqrt{N_B}$, where $N_{MC}$ and 
$N_B$ are the numbers of MC and 
background events, respectively, after all selection criteria were 
applied. The data within the signal windows were unmasked only after 
this optimization. We used asymmetric windows for the decay modes 
containing electrons to allow for the bremsstrahlung low-energy tail. 
The signal windows were: 
$1.83<M(D^0)<1.90$ GeV/$c^2$ for $\mu \mu $ and 
$1.76<M(D^0)<1.90$ GeV/$c^2$ for $ee$ and $\mu e$ modes.

The upper limit for each branching fraction $B_{X}$ was calculated 
using the following formula: 
\begin{equation}
B_{X}=\frac{N_{X}}{N_{\mathrm{Norm}}}
\frac{\varepsilon _{\mathrm{Norm}}}{\varepsilon _{X}}
\times B_{\mathrm{Norm}};~\mathrm{where}~ 
\frac{\varepsilon _{\mathrm{Norm}}}{\varepsilon _{X}}=
\frac{f_{\mathrm{Norm}}^{\mathrm{MC}}}{f_{X}^{\mathrm{MC}}}.
\label{BReqn}
\end{equation}
$N_{X}$ is the 90$\%$ confidence level (CL) upper limit on the 
number of decays for the rare or forbidden decay mode $X$ 
and $B_{\mathrm{Norm}}$ is the normalization mode branching 
fraction obtained from the Particle Data Group \cite{PDG}. 
$\varepsilon_{\mathrm{Norm}}$ and $\varepsilon_{X}$ are the detection 
efficiencies while 
$f_{\mathrm{Norm}}^{\mathrm{MC}}$ and $f_{X}^{\mathrm{MC}}$ are 
the fractions of Monte Carlo events that were reconstructed and passed 
the final selection criteria, for the normalization and decay modes, 
respectively. 

The 90$\%$ CL upper limits $N_{X}$ are calculated using the method of 
Feldman and Cousins \cite{Feldman} to account for background, and then 
corrected for systematic errors by the method of Cousins and Highland 
\cite{Cousins}. In these methods, the numbers of signal events are 
determined by simple counting, not by a fit. Upper limits are 
determined using the number of candidate events observed and 
expected number of background events within the 
signal region. (See Refs. \cite{FCNCnew,4prong} for a more detailed 
discussion of backgrounds.)
\section{The $D^+\rightarrow h\ell \ell $, 
$D_{s}^+\rightarrow h\ell \ell $ and $D^0\rightarrow \ell^+\ell^-$ 
Analysis}
\vskip -10pt
We normalized the sensitivity of our search to topologically similar 
Cabibbo-favored decays. For the $D^{+}$ decays we used 
$24010\pm 166 ~D^+\rightarrow K^-\pi^+\pi^+$; for $D_{s}^{+}$ 
decays we used $782\pm 30 ~D_{s}^{+}\rightarrow \phi \pi^+$; and for 
$D^{0}$ decays we used $25210\pm 179 ~D^0\rightarrow K^-\pi^+$ events. 
The widths of our normalization modes were 10.5 MeV/$c^{\,2}$ for 
$D^{+}$, 9.5 MeV/$c^{\,2}$ for $D_{s}^{+}$, and 12 MeV/$c^{\,2}$ for 
$D^{0}$. The results are shown in Table \ref{Results1} and compared 
with previous results in Figure \ref{BR1}.
\vskip -5pt
\begin{table}[h!]
\caption{
E791 90$\%$ confidence level branching fraction upper limits for 
$D^+\rightarrow h\ell \ell $, $D_{s}^+\rightarrow h\ell \ell $ and 
$D^0\rightarrow \ell^+\ell^-$.}
\label{Results1}
\vskip 5pt
\begin{center}
\begin{tabular}{|ll|ll|ll|} \hline
Mode&E791 Limit&Mode&E791 Limit&Mode&E791 Limit \\
\hline
\vspace*{-10pt} & & & & & \\
 $D^{+}\rightarrow \pi ^{+}\mu ^{+}\mu ^{-}$&$1.5\times 10^{-5}$&
 $D^{+}\rightarrow \pi ^{+}e^{+}e^{-}$&$5.2\times 10^{-5}$&
 $D^{+}\rightarrow \pi ^{+}\mu ^{\pm }e^{\mp }$&$3.4\times 10^{-5}$\\
 $D^{+}\rightarrow \pi ^{-}\mu ^{+}\mu ^{+}$&$1.7\times 10^{-5}$&
 $D^{+}\rightarrow \pi ^{-}e^{+}e^{+}$&$9.6\times 10^{-5}$&
 $D^{+}\rightarrow \pi ^{-}\mu ^{+}e^{+}$&$5.0\times 10^{-5}$\\
 $D^{+}\rightarrow K^{+}\mu ^{+}\mu ^{-}$&$4.4\times 10^{-5}$&
 $D^{+}\rightarrow K^{+}e^{+}e^{-}$&$2.0\times 10^{-4}$&
 $D^{+}\rightarrow K^{+}\mu ^{\pm }e^{\mp }$&$6.8\times 10^{-5}$\\
\hline
\vspace*{-10pt} & & & & & \\
 $D_{s}^{+}\rightarrow K^{+}\mu ^{+}\mu ^{-}$&$1.4\times 10^{-4}$&
 $D_{s}^{+}\rightarrow K^{+}e^{+}e^{-}$&$1.6\times 10^{-3}$&
 $D_{s}^{+}\rightarrow K^{+}\mu ^{\pm }e^{\mp }$&$6.3\times 10^{-4}$\\
 $D_{s}^{+}\rightarrow K^{-}\mu ^{+}\mu ^{+}$&$1.8\times 10^{-4}$&
 $D_{s}^{+}\rightarrow K^{-}e^{+}e^{+}$&$6.3\times 10^{-4}$&
 $D_{s}^{+}\rightarrow K^{-}\mu ^{+}e^{+}$&$6.8\times 10^{-4}$\\
 $D_{s}^{+}\rightarrow \pi ^{+}\mu ^{+}\mu ^{-}$&$1.4\times 10^{-4}$&
 $D_{s}^{+}\rightarrow \pi ^{+}e^{+}e^{-}$&$2.7\times 10^{-4}$&
 $D_{s}^{+}\rightarrow \pi ^{+}\mu ^{\pm }e^{\mp }$&$6.1\times 10^{-4}$\\
 $D_{s}^{+}\rightarrow \pi ^{-}\mu ^{+}\mu ^{+}$&$8.2\times 10^{-5}$&
 $D_{s}^{+}\rightarrow \pi ^{-}e^{+}e^{+}$&$6.9\times 10^{-4}$&
 $D_{s}^{+}\rightarrow \pi ^{-}\mu ^{+}e^{+}$&$7.3\times 10^{-4}$\\
\hline
\vspace*{-10pt} & & & & & \\
 $D^{0}\rightarrow \mu ^{+}\mu ^{-}$&$5.2\times 10^{-6}$&
 $D^{0}\rightarrow e^{+}e^{-}$&$6.2\times 10^{-6}$&
 $D^{0}\rightarrow \mu ^{\pm }e^{\mp }$&$8.1\times 10^{-6}$\\
\hline
\end{tabular}
\end{center}
\end{table}
\section{The $D^0\rightarrow V\ell^+\ell^-$ and 
$D^0\rightarrow hh\ell \ell$ Analysis}
\vskip -10pt
There were a few minor differences between this analysis and our 
previous analysis as discussed above. First, we examined resonant 
modes, where the mass ranges used were: 
$\left| m_{\pi ^+\pi ^-} - m_{\rho ^0}\right| <150$ MeV/$c^2$, 
$\left| m_{K^-\pi ^+} - m_{\Kstar}\right| <55$ MeV/$c^2$, and 
$\left| m_{K^+K^-} - m_{\phi}\right| <10$ MeV/$c^2$.
We normalized the sensitivity of each search to 
similar hadronic 3-body (resonant) or 4-body (non-resonant) decays. 
One exception is the case of 
$D^{0}\rightarrow \rho^{0} \ell ^{\pm }\ell ^{\mp }$ where we 
normalize to nonresonant  
$D^{0}\rightarrow \pi ^+\pi ^-\pi ^+\pi ^-$ because no published 
branching fraction exists for 
$D^0\rightarrow \rho ^{0}\pi ^+\pi ^-$. Table \ref{Norm} lists the 
normalization mode used for each signal mode and 
the fitted numbers of normalization data events ($N_{\mathrm{Norm}}$).
\begin{table}[ht!]
\caption[]{
Normalization modes used for $D^0\rightarrow V\ell^+\ell^-$ and 
$D^0\rightarrow hh\ell \ell$.}
\label{Norm}
\vskip 5pt
\tabcolsep=4.0pt
\begin{center}
\begin{tabular}{|lll|lll|} 
\hline
Decay Mode&Norm. Mode&$N_{\mathrm{Norm}}$&
Decay Mode&Norm. Mode&$N_{\mathrm{Norm}}$\\
\hline
\vspace*{-10pt} & & & & & \\
$D^{0}\rightarrow \rho^{0} \ell ^{\pm }\ell ^{\mp }$&
$D^{0}\rightarrow \pi ^+\pi ^-\pi ^+\pi ^-$& 2049$\pm$53& 
$D^{0}\rightarrow \Kstar \ell ^{\pm }\ell ^{\mp }$&
$D^{0}\rightarrow \Kstar \pi ^+\pi ^-$& 5451$\pm$72\\ 
$D^{0}\rightarrow \phi \ell ^{\pm }\ell ^{\mp }$&
$D^{0}\rightarrow \phi \pi ^+\pi ^-$& 113$\pm$19&  
$D^{0}\rightarrow \pi \pi \ell\ell $& 
$D^{0}\rightarrow \pi ^+\pi ^-\pi ^+\pi ^-$& 2049$\pm$53\\
$D^{0}\rightarrow K\pi \ell\ell $&
$D^{0}\rightarrow K^-\pi ^+\pi ^-\pi ^+$&11550$\pm$113& 
$D^{0}\rightarrow KK\ell\ell $&
$D^{0}\rightarrow K^+K^-\pi ^+\pi ^-$& 406$\pm$41\\
\hline
\end{tabular}
\end{center}
\end{table}

\begin{figure}[hb!]
   \begin{minipage}{3.15in}
    \centerline{\epsfxsize 3.15 truein \epsfbox{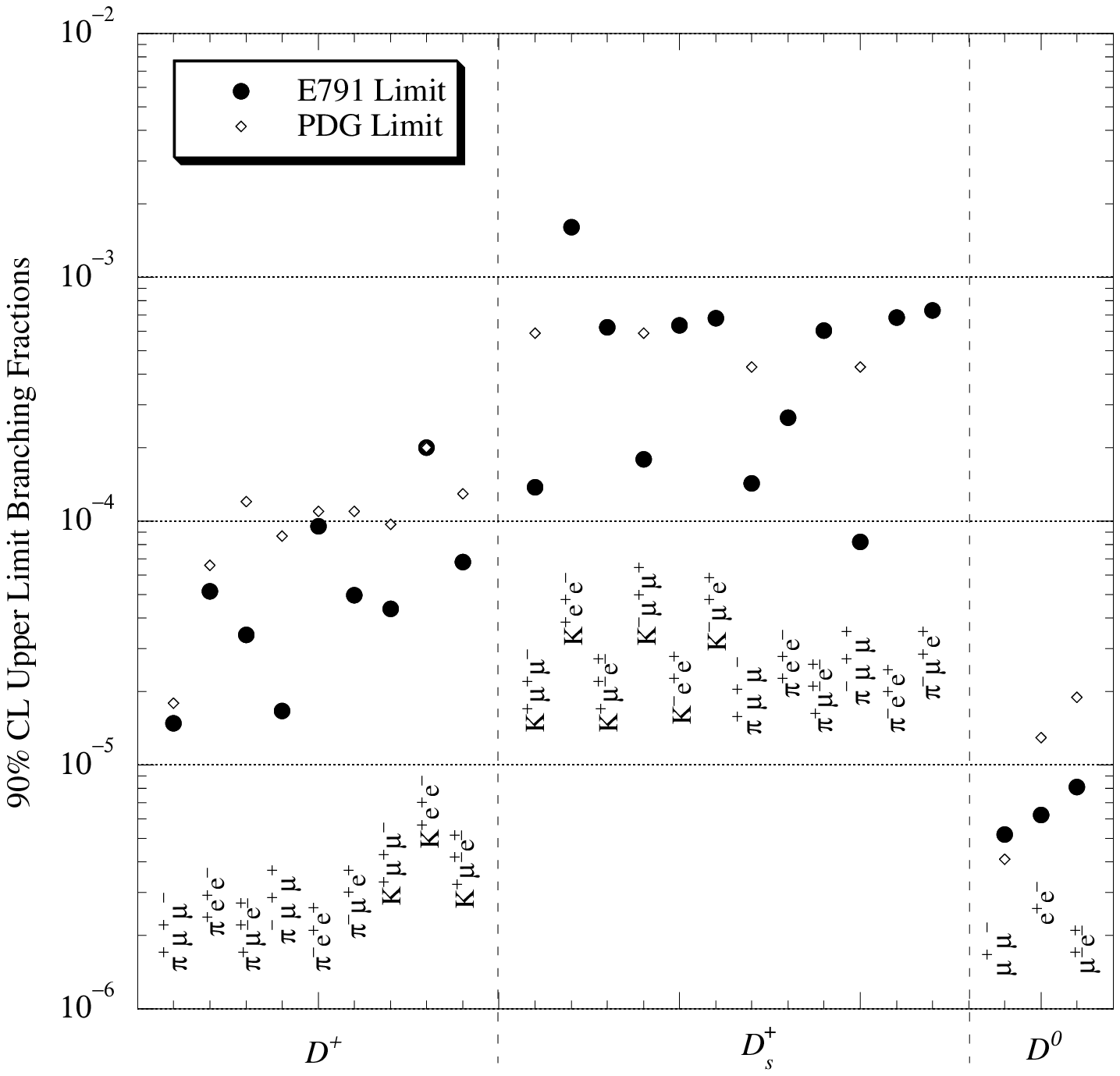}}
    \caption{
\small 
Comparison of the 90$\%$ CL upper-limit branching fractions 
from E791 data (dark circles) with existing limits (open diamonds) from 
the 1998 PDG$^{14}$.
}
    \label{BR1}
   \end{minipage}
\hspace{0.4in}
   \begin{minipage}{3.05in}
    \centerline{\epsfxsize 3.05 truein \epsfbox{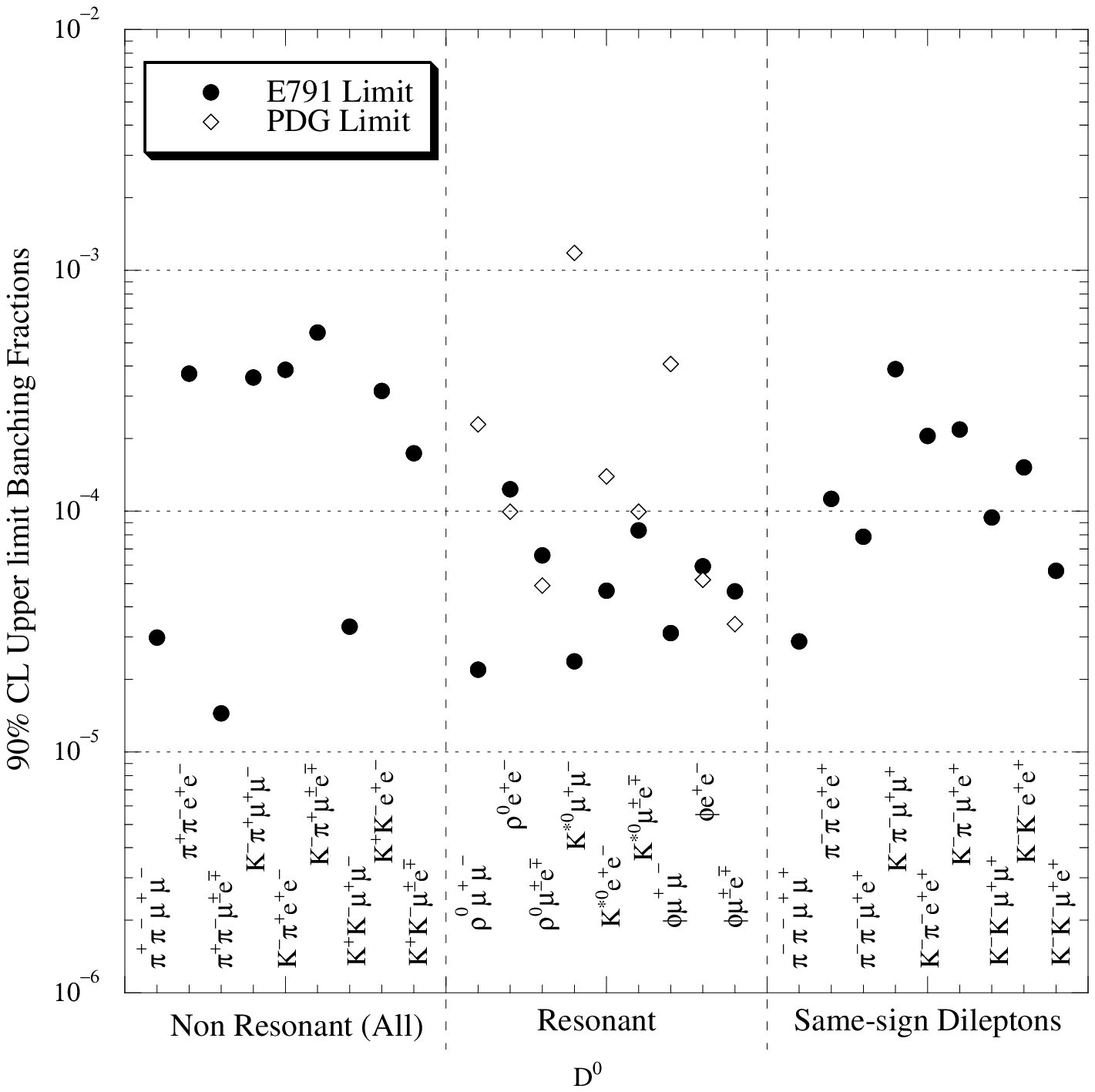}}
    \caption{
\small Comparison of the 90$\%$ CL upper-limit branching fractions 
from E791 data (dark circles) with existing limits (open diamonds) from 
the 2000 PDG$^{14}$. 
}
    \label{BR2}
   \end{minipage}
\end{figure}

\noindent
The final results are shown in Table \ref{Results2} and compared 
with previous results in Figure \ref{BR2}.
\begin{table}[ht!]
\caption[]{
E791 90$\%$ confidence level branching fraction upper limits for 
$D^0\rightarrow V\ell^+\ell^-$ and $D^0\rightarrow hh\ell \ell$.}
\label{Results2}
\vskip -10pt
\tabcolsep=4.0pt
\begin{center}
\begin{tabular}{|ll|ll|ll|}
\hline
Mode&E791 Limit&Mode&E791 Limit&Mode&E791 Limit\\
\hline
\vspace*{-10pt} & & & & & \\
 $D^{0}\rightarrow \pi ^{+}\pi ^{-}\mu ^{+}\mu ^{-}$&$3.0\times 10^{-5}$&
 $D^{0}\rightarrow \pi ^{+}\pi ^{-}e^{+}e^{-}$&$3.7\times 10^{-4}$&
 $D^{0}\rightarrow \pi ^{+}\pi ^{-}\mu ^{\pm }e^{\mp }$&$1.5\times 10^{-5}$\\
 $D^{0}\rightarrow K^{-}\pi ^{+}\mu ^{+}\mu ^{-}$&$3.6\times 10^{-4}$&
 $D^{0}\rightarrow K^{-}\pi ^{+}e^{+}e^{-}$&$3.9\times 10^{-4}$&
 $D^{0}\rightarrow K^{-}\pi ^{+}\mu ^{\pm }e^{\mp }$&$5.5\times 10^{-4}$\\
 $D^{0}\rightarrow K^{+}K^{-}\mu ^{+}\mu ^{-}$&$3.3\times 10^{-5}$&
 $D^{0}\rightarrow K^{+}K^{-}e^{+}e^{-}$&$3.2\times 10^{-4}$&
 $D^{0}\rightarrow K^{+}K^{-}\mu ^{\pm }e^{\mp }$&$1.8\times 10^{-4}$\\
\hline
\vspace*{-10pt} & & & & &\\
 $D^{0}\rightarrow \rho ^{0}\mu ^{+}\mu ^{-}$&$2.2\times 10^{-5}$&
 $D^{0}\rightarrow \rho ^{0}e^{+}e^{-}$&$1.2\times 10^{-4}$&
 $D^{0}\rightarrow \rho ^{0}\mu ^{\pm }e^{\mp }$&$6.6\times 10^{-5}$\\
 $D^{0}\rightarrow \Kstar \mu ^{+}\mu ^{-}$&$2.4\times 10^{-5}$&
 $D^{0}\rightarrow \Kstar e^{+}e^{-}$&$4.7\times 10^{-5}$&
 $D^{0}\rightarrow \Kstar \mu ^{\pm }e^{\mp }$&$8.3\times 10^{-5}$\\
 $D^{0}\rightarrow \phi \mu ^{+}\mu ^{-}$&$3.1\times 10^{-5}$&
 $D^{0}\rightarrow \phi e^{+}e^{-}$&$5.9\times 10^{-5}$&
 $D^{0}\rightarrow \phi \mu ^{\pm }e^{\mp }$&$4.7\times 10^{-5}$\\
\hline
\vspace*{-10pt} & & & & &\\
 $D^{0}\rightarrow \pi ^{-}\pi ^{-}\mu ^{+}\mu ^{+}$&$2.9\times 10^{-5}$&
 $D^{0}\rightarrow \pi ^{-}\pi ^{-}e^{+}e^{+}$&$1.1\times 10^{-4}$&
 $D^{0}\rightarrow \pi ^{-}\pi ^{-}\mu ^{+}e^{+}$&$7.9\times 10^{-5}$\\
 $D^{0}\rightarrow K^{-}\pi ^{-}\mu ^{+}\mu ^{+}$&$3.9\times 10^{-4}$&
 $D^{0}\rightarrow K^{-}\pi ^{-}e^{+}e^{+}$&$2.1\times 10^{-4}$&
 $D^{0}\rightarrow K^{-}\pi ^{-}\mu ^{+}e^{+}$&$2.2\times 10^{-4}$\\
 $D^{0}\rightarrow K^{-}K^{-}\mu ^{+}\mu ^{+}$&$9.4\times 10^{-5}$&
 $D^{0}\rightarrow K^{-}K^{-}e^{+}e^{+}$&$1.5\times 10^{-4}$&
 $D^{0}\rightarrow K^{-}K^{-}\mu ^{+}e^{+}$&$5.7\times 10^{-5}$\\
\hline
\end{tabular}
\end{center}
\end{table}
\section{Conclusion}
\vskip -10pt
We used a ``blind'' analysis of data from Fermilab 
experiment E791 to obtain upper limits on the dilepton branching 
fractions for 51 flavor-changing neutral current, lepton-number 
violating, and lepton-family violating decays of $D^+$, $D_{s}^{+}$, 
and $D^0$ mesons. No evidence for any of these 2, 3 and 4-body decays 
was found. Therefore, we presented upper limits on the branching 
fractions at the 90$\%$ confidence level. Eighteen limits represented 
significant improvements over previously published results. Twenty-six 
of the remaining modes had no previously reported limits.
\section*{Acknowledgments}
\vskip -10pt
We thank the staffs of Fermilab and 
participating institutions. This research was supported by 
the Brazilian Conselho Nacional de Desenvolvimento Cient\'\i fico e 
Tecnol\'ogico, CONACyT (Mexico), the Israeli Academy of Sciences and 
Humanities, the U.S.-Israel Binational 
Science Foundation, and the U.S.~National Science Foundation 
and Department~of Energy. 
The Universities Research Association~operates Fermilab for 
the U.S.~Department~of Energy.

\end{document}